\documentclass[12pt]{article}

\usepackage{array} 
\usepackage{epsfig}
\usepackage{amssymb}
\usepackage{amsmath}                          
\usepackage{graphics,graphpap}

\renewcommand{\thefootnote}{\fnsymbol{footnote}}

\newcommand{\qz}{(q z)}
\newcommand{\ub}{\bar u}
\newcommand{\quark}{\langle \bar q q\rangle}
\newcommand{\mixed}{\langle \bar q \sigma gG q\rangle}
\newcommand{\squark}{\langle \bar s s\rangle}
\newcommand{\smixed}{\langle \bar s \sigma gG s\rangle}
\newcommand{\gluon}{\left\langle \frac{\alpha_s}{\pi}\,G^2\right\rangle}

\begin{document}

\begin{titlepage}
\begin{flushright}\begin{tabular}{l}
IPPP/05/57\\
DCPT/05/114
\end{tabular}
\end{flushright}
\vskip1.5cm
\begin{center}
   {\Large \bf \boldmath SU(3) Breaking of Leading--Twist\\[5pt]
K and K$^*$  Distribution Amplitudes -- a Reprise}
    \vskip1.3cm {\sc
Patricia Ball\footnote{Patricia.Ball@durham.ac.uk} and Roman 
Zwicky\footnote{Roman.Zwicky@durham.ac.uk}
  \vskip0.5cm
        {\em IPPP, Department of Physics, 
University of Durham, Durham DH1 3LE, UK}} \\
\vskip2.5cm 


\vskip3cm

{\large\bf Abstract:\\[10pt]} \parbox[t]{\textwidth}{
We review the status of the leptonic decay constants $f_K$ and
$f_K^{\parallel,\perp}$ of the  $K$ and $K^*$, respectively, and the 
SU(3) breaking quantities $a_1(K)$ and
$a_{1}^{\parallel,\perp}(K^{*})$, 
the first Gegenbauer-moments of the leading-twist
distribution amplitudes of $K$ and $K^*$. We obtain new
predictions from QCD sum rules which are relevant for
the calculation of $K$ and $K^*$ form factors, for instance $T_1^{B\to
  K*}$, which determines the decay $B\to K^*\gamma$, and for QCD
factorisation calculations of nonleptonic $B$ decays into strange
mesons, for instance $B\to K\pi$.}

\vfill

{\em submitted to Physics Letters B}
\end{center}
\end{titlepage}

\setcounter{footnote}{0}
\renewcommand{\thefootnote}{\arabic{footnote}}

\newpage

\section{Introduction}\label{sec:1}

Hadronic light-cone distribution amplitudes (DAs) of 
leading twist play an essential
r\^{o}le in the QCD description of hard exclusive processes. 
DAs enter the amplitudes of processes to which collinear factorisation
theorems apply and were first discussed in the seminal
papers by Brodsky, Lepage and others \cite{exclusive}. More recently,
collinear factorisation has been shown to apply, to leading order in
an expansion in $1/m_b$, also to a large class
of nonleptonic B decays \cite{BBNS}, which has opened a new and
exciting area of
applications of meson DAs. These decays, and in particular
their CP asymmetries, are currently being studied at the B factories
BaBar and Belle and are expected to yield essential information about the
pattern of CP violation and potential sources of flavour violation
beyond the SM. 
The aim of this letter is to provide a reanalysis of 
SU(3) breaking effects in leading-twist $K$ and $K^*$ DAs, using QCD
sum rules.
Our letter is both a sequence to and an extension of previous work 
reported in Refs.~\cite{Russians,CZreport,elena,alex,lenz}. The
results are of immediate relevance for all predictions of $B\to
(K,K^*)$ decay processes calculated in QCD factorisation. 

 We define two-particle DAs as matrix elements of
  quark-antiquark gauge-invariant nonlocal operators at light-like
  separations $z_\mu$ with $z^2=0$. For definiteness
  we consider distributions of mesons with an $s$ quark and a light
  antiquark $\bar q$. To leading-twist accuracy, the complete set of 
  distributions comprises three DAs (we use
  the notation $\hat{z} = z^\mu\gamma_\mu$ for arbitrary four-vectors $z$):
  \begin{eqnarray}
    \langle 0 |\bar q(z)\hat{z}\gamma_5 [z,0]s(0)
  |K(q)\rangle &=& i  f_K \qz 
  \int_0^1 du\, e^{-i\ub\qz} \phi_K(u)\,,\nonumber\\
  \langle 0 |\bar q(z)\hat{z}[z,0] s(0)
  |K^{*}(q,\lambda)\rangle &=& (e^{(\lambda)} z)
  f_K^\parallel m_{K^*}\int_0^1 du\, e^{-i\ub\qz}
  \phi_K^\parallel(u),\nonumber\\
  \langle 0 |\bar q(z)\sigma_{\mu\nu}[z,0]s(0)
  |K^{*}(q,\lambda)\rangle & = &
  i(e^{(\lambda)}_\mu q_\nu -e^{(\lambda)}_\nu q_\mu)
  f_K^\perp(\mu) \int_0^1 du\, e^{-i\ub\qz} \phi_K^\perp(u),
  \label{eq:defDAs}
  \end{eqnarray}
  with the Wilson-line
  $$
  [z,0] = \mbox{Pexp}\,\left[ig\int_0^1 d\alpha\, z^\mu A_\mu(\alpha
    z)\right]
  $$
inserted between quark fields to render the matrix elements
gauge-invariant. 
  In the above definitions, $e^{(\lambda)}_\nu$ is the
   polarization vector of a vector meson with polarisation
   $\lambda$; there are two leading-twist 
DAs for vector mesons, $\phi_K^\parallel$
   and $\phi_K^\perp$, corresponding to
   longitudinal and transverse polarisation, respectively.
 The integration variable $u$ is the (longitudinal)
meson momentum fraction carried by the quark, $\ub \equiv 1-u$ the
  momentum fraction carried by the antiquark.  The normalisation constants
  $f_K^{(\parallel,\perp)}$ are defined by the local limit of
  Eqs.~(\ref{eq:defDAs}) and chosen in such a way that 
  \begin{equation}\label{eq:norm} \int_0^1 du\, \phi(u)=1\end{equation}
  for all three distributions $\phi_K,
  \phi_K^\parallel , \phi_K^\perp$. 

The most relevant parameters characterising SU(3) breaking are the
decay constants $f_K$ and $f_{K}^{\perp,\parallel}$, and the first 
(Gegenbauer) moments 
of the corresponding leading-twist DAs, $a_1(K)$ and
$a_1^{\perp,\parallel}(K^*)$, defined as
\begin{equation}
a_1(K) = \frac{5}{3}\, \int_0^1 du\,
 (2u-1)\phi_K(u).
\end{equation}
and correspondingly for $a_1^{\parallel,\perp}(K^*)$.
$a_1$ describes the
difference of the average longitudinal momenta of the quark and
antiquark in the two-particle Fock-state component of the meson, a quantity
that vanishes for particles with equal-mass quarks (particles with 
definite G-parity). The decay constants $f_K$ and
$f_{K}^\parallel$ can be extracted from experiment;
$f_{K}^\perp$ has been calculated both from lattice \cite{lattbec}
and from QCD sum rules, e.g.\ Ref.~\cite{BZvector}. Although the
results are in mutual agreement, the QCD sum rule calculations are
still subject to improvement as we shall discuss in this letter. The 
situation is much less clear with $a_1$: no lattice
calculation of this quantity has been attempted yet, so essentially
all available information on $a_1$ comes from QCD sum rule
calculations. To date, three different types of sum rules have been used to 
calculate $a_1$:
\begin{itemize}
\item sum rules based on the correlation function of two biquark
  currents of equal chirality, so-called {\em diagonal} sum rules 
\cite{Russians,CZreport,alex};
\item  sum rules based on the correlation function of two biquark
  currents of different chirality, so-called {\em nondiagonal} sum rules 
\cite{Russians,CZreport,elena};
\item exact operator identities relating $a_1(K)$ and
  $a_1^\parallel(K^*)$ to quark-quark-gluon matrix elements of the
  meson, which in turn are calculated from QCD sum rules \cite{lenz}.
\end{itemize}
The basic argument in favour of nondiagonal sum rules is that they are
of first order in SU(3)-breaking quantities: the perturbative
contribution is $\sim O(m_s)$, the leading nonperturbative
contribution $\sim \langle \bar q q\rangle - \langle \bar s s
\rangle$, whereas for diagonal correlation functions the corresponding
contributions are $\sim
O(m_s^2)$ and $m_s\langle \bar s s \rangle - m_u \langle \bar u u
\rangle$, respectively.
The original calculation of Chernyak and Zhitnitsky using a
nondiagonal sum rule yielded $a_1\sim 0.1$ \cite{Russians,CZreport}, but
unfortunately 
suffers from a sign-mistake in the perturbative contribution. This
mistake 
was corrected in Ref.~\cite{elena}, which however entails that
the two leading contributions come with different sign and cancel to a
large extent. As pointed out in Ref.~\cite{alex,lenz}, the resulting sum
rules are sensitive to poorly constrained higher-order perturbative
and nonperturbative corrrections and hence numerically unreliable. 
Alternative sum rules for $a_1$ 
come from diagonal correlation functions, a route that was
followed, for $a_1(K)$, in Ref.~\cite{alex}. Yet another possibility
to pin down the elusive $a_1$ is offered by exploiting exact 
operator identities that relate $a_1$ to quark-quark-gluon 
matrix-elements which in turn are calculated by QCD sum
rules; the corresponding results for $a_1(K)$ and $a_1^\parallel(K^*)$
can be found in Ref.~\cite{lenz}. 
In this letter we derive and analyse 
diagonal sum rules for the decay constants $f_K$ and all three
Gegenbauer moments $a_1(K)$ and $a_1^{\parallel,\perp}(K^*)$. We plan
to come back to the analysis of operator identities and the
corresponding sum rules for quark-quark-gluon matrix elements in a
future publication.

Our letter is organised as follows: in Sec.~\ref{sec:2} we calculate
and analyse QCD sum rules for $f_K$, $f_K^{\parallel,\perp}$ and
$a_1(K)$, $a_1^{\parallel,\perp}(K^*)$. 
We summarise and conclude in Sec.~\ref{sec:4}. The
appendix contains some remarks about the calculation of diagonal sum
rules.
 
\section{\boldmath QCD Sum Rules for $f_K^{(\parallel,\perp)}$ and 
$a_1^{(\parallel,\perp)}(K^{(*)})$}\label{sec:2}

QCD sum rules are an established method for the
calculation of ha\-dro\-nic matrix elements, see
Refs.~\cite{SVZ,colalex} for the original papers and a recent review. 
The key feature
of the method is the use of analyticity to relate the local
short-distance operator
product expansion (OPE) of a correlation function of two currents,
\begin{equation}\label{eq1}
\Pi = i\int d^4y e^{iqy} \langle 0 | T J_1(y) J_2(0) | 0 \rangle =
\sum_n C_n(q^2) \langle O_n\rangle\equiv \Pi^{\mbox{\scriptsize OPE}}
\end{equation}
around $y = 0$ (as opposed to a
light-cone expansion aound $y^2=0$, which is appropriate
for form factor calculations, cf.~\cite{FFs})
valid for $Q^2\equiv -q^2 \ll 0$, to its dispersion relation in terms
of hadronic contributions,
\begin{equation}\label{eq2}
\Pi = \int_0^\infty ds\, \frac{\rho(s)}{s-q^2-i0}\equiv
\Pi^{\mbox{\scriptsize had}},
\end{equation}
where $\rho(s)$ is the spectral density of the correlation function
along its physical cut. 
The OPE yields a series of local operators of increasing dimension
whose expectation values $\langle O_n\rangle$ 
in the nonperturbative (physical) vacuum are
the so-called condensates. In the sum rules analysed in this letter, we
take into account the condensates and parameters listed in Tab.~\ref{tab:cond}.
As for the strange quark mass, we would like to recall that with
present data there is a hint of a discrepancy between unquenched
$n_f=2$ and $n_f=2+1$ results, the latter ones favouring smaller values
$\overline{m}_s(2\,{\rm GeV}) = (78\pm 10)\,{\rm MeV}$
\cite{Wittig}. Awaiting the clarification of this situation, we choose
to stay with the result from  $n_f=2$ flavours given in
Tab.~\ref{tab:cond}. 

\begin{table}[b]
\renewcommand{\arraystretch}{1.3}
\addtolength{\arraycolsep}{3pt}
$$
\begin{array}{|r@{\:=\:}l||r@{\:=\:}l|}
\hline 
\quark & (-0.24\pm0.01)^3\,\mbox{GeV}^3 & \squark & (0.8\pm0.1)\,\quark\\
\mixed & (0.8\pm0.1)\,\mbox{GeV}^2\:\quark &  \smixed & (0.8\pm0.1)
\mixed\\[6pt]
\displaystyle \gluon & 0.012\, {\rm GeV}^4 & \multicolumn{2}{l|}{}\\[6pt]\hline
\multicolumn{4}{|l|}{\overline{m}_s(2\,\mbox{GeV}) = (100\pm
20)\,\mbox{MeV}~\longleftrightarrow~ \overline{m}_s(1\,\mbox{GeV})
= (137\pm 27)\,\mbox{MeV}}\\
\multicolumn{4}{|l|}{\hskip2cm\alpha_s(1\,\mbox{GeV}) = 0.534 
~\longleftrightarrow~
\Lambda_{\mbox{\scriptsize QCD}}^{(3)\mbox{\scriptsize NLO}} = 
384\,\mbox{MeV}}\\\hline
\end{array}
$$
\renewcommand{\arraystretch}{1}
\addtolength{\arraycolsep}{-3pt}
\vskip-10pt
\caption[]{Input parameters for sum rules at the
  renormalisation scale $\mu=1\,$GeV. The value of $m_s$ is obtained
  from 
  unquenched lattice calculations with $n_f=2$ flavours 
as summarised in \cite{Wittig}.
}\label{tab:cond}
\end{table}
The representation of the correlation function in terms of hadronic
matrix elements can be written as
$$
\rho(s) = f \delta(s-m_M^2) + \rho^{\mbox{\scriptsize cont}}(s),
$$
where $m_M$ is the mass of the lowest-lying state coupling to the currents
$J_{1,2}$ and $\rho^{\mbox{\scriptsize cont}}$ pa\-ra\-me\-trises all
contributions to the correlation function apart from the
ground state. $f$, the residue of the ground state pole, is the
quantity to be determined. A QCD sum rule that allows one to do
so is obtained by equating the representations (\ref{eq1}) and
(\ref{eq2}) and
implementing the following (model) assumptions:
\begin{itemize}
\item $\rho^{\mbox{\scriptsize cont}}$ is approximated by the
   spectral density obtained from the OPE above a certain threshold,
   i.e.\ 
  $\rho^{\mbox{\scriptsize cont}}\to
\rho^{\mbox{\scriptsize OPE}}(s) \theta(s-s_0)$ with 
   $s_0\approx (m_M+\Delta)^2$ being the continuum threshold, where 
$\Delta\sim O(\Lambda_{\mbox{\scriptsize QCD}})$ 
is an excitation
  energy to be determined within the method. This assumption relies on
  the validity of semiglobal quark-hadron duality;
\item instead of the weight-functions $1/(q^2)^n$ and $1/(s-q^2)$, one
   uses different weight-functions which are optimised to
   (exponentially) suppress 
  effects of $\rho(s)$ for large values of $s$ and at the same time
   also suppress higher-dimensional condensates by factorials. This is
   achieved by Borel transforming the correlation function: 
   ${\cal B}\,1/(s-q^2) = 1/M^2\exp(-s/M^2)$. A window of viable
   values of the Borel parameter $M^2$ and the continuum threshold
   $s_0$ has to be determined within the
   method itself by looking for a maximum region of minimum sensitivity (a
   plateau) in both $M^2$ and $s_0$;
\item the OPE of $\Pi$ can be truncated after a few terms. As 
   is well known, this condition is fulfilled only for
  low moments, whereas for higher moments of the DAs in $(u-\bar u)$
   the nonperturbative terms become dominant. 
\end{itemize}
After subtraction of the integral over $\rho^{\mbox{\scriptsize
    OPE}}$ above $s_0$ from both sides, the final sum rule reads
\begin{equation}\label{eq:SR}
{\cal B}_{\mbox{\scriptsize sub}}\,\Pi^{\mbox{\scriptsize OPE}} \equiv
 \frac{1}{M^2}\int_0^{s_0} ds\,e^{-s/M^2}\,
\rho^{\mbox{\scriptsize OPE}}(s) = \frac{f}{M^2}\,e^{-m_M^2/M^2},
\end{equation}
which gives the hadronic quantity $f$ as a function of the Borel
parameter $M^2$ and the continuum threshold $s_0$ (and the condensates
and short-distance parameters from the OPE).

\begin{figure}[p]
$$\epsfysize=0.19\textheight\epsffile{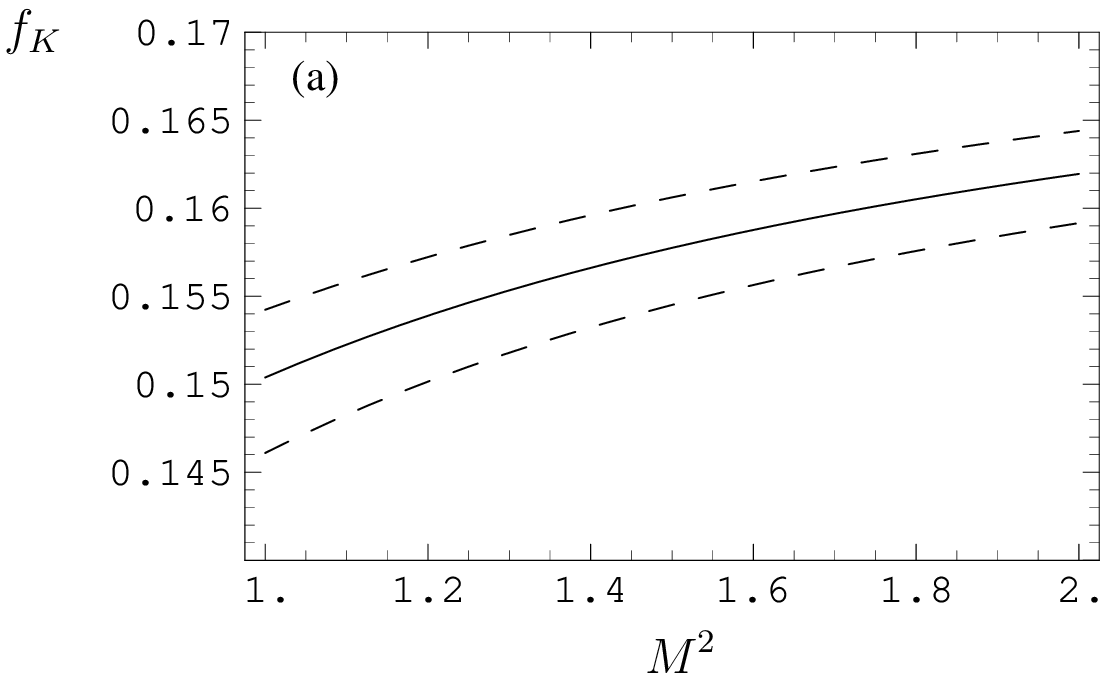}\qquad
\epsfysize=0.19\textheight\epsffile{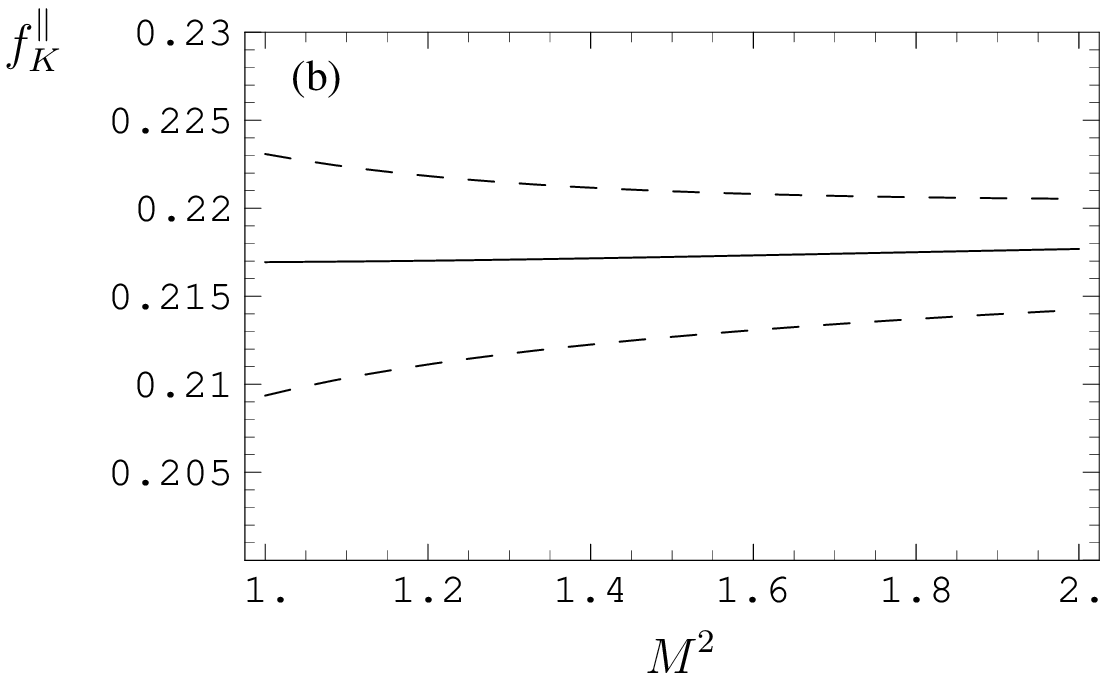}$$
\vskip-10pt
\caption[]{(a) $f_K$ as function of the Borel parameter
  $M^2$ for $s_0=1.1\,{\rm GeV}^2$. Solid line: central values of
  input parameters, dashed lines: variation of $f_K$ within the
  allowed range of input parameters. (b): the same for
  $f_K^\parallel$ with $s_0=1.7\,{\rm GeV}^2$.}\label{fig:1}
\vskip-10pt
$$\epsfysize=0.19\textheight\epsffile{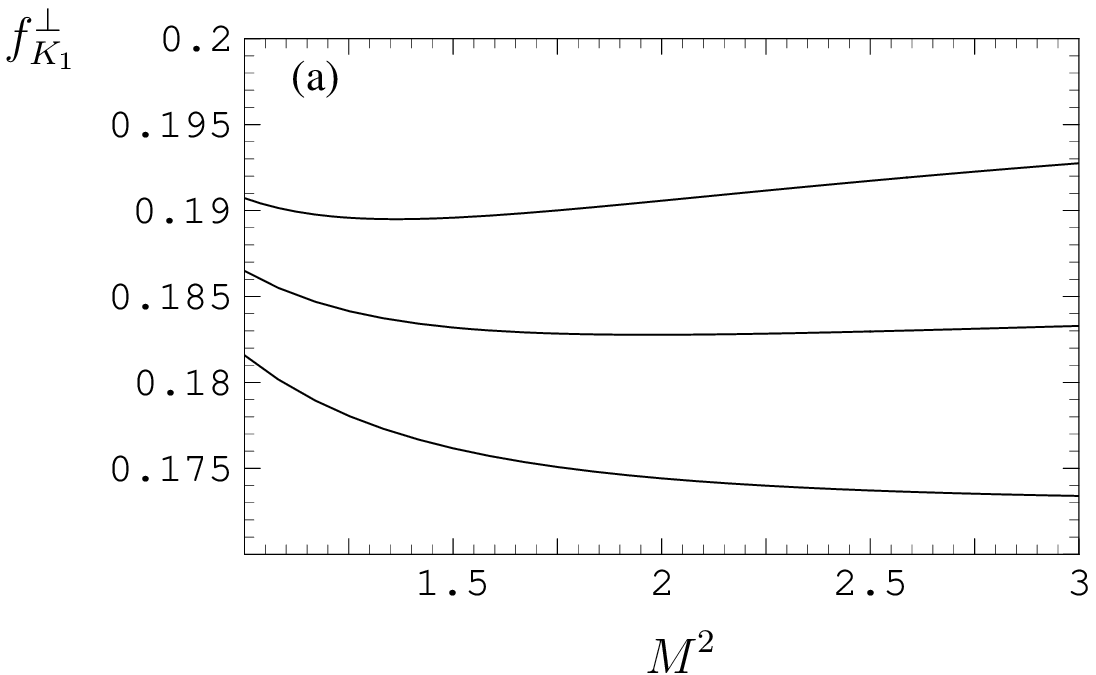}\qquad
\epsfysize=0.19\textheight\epsffile{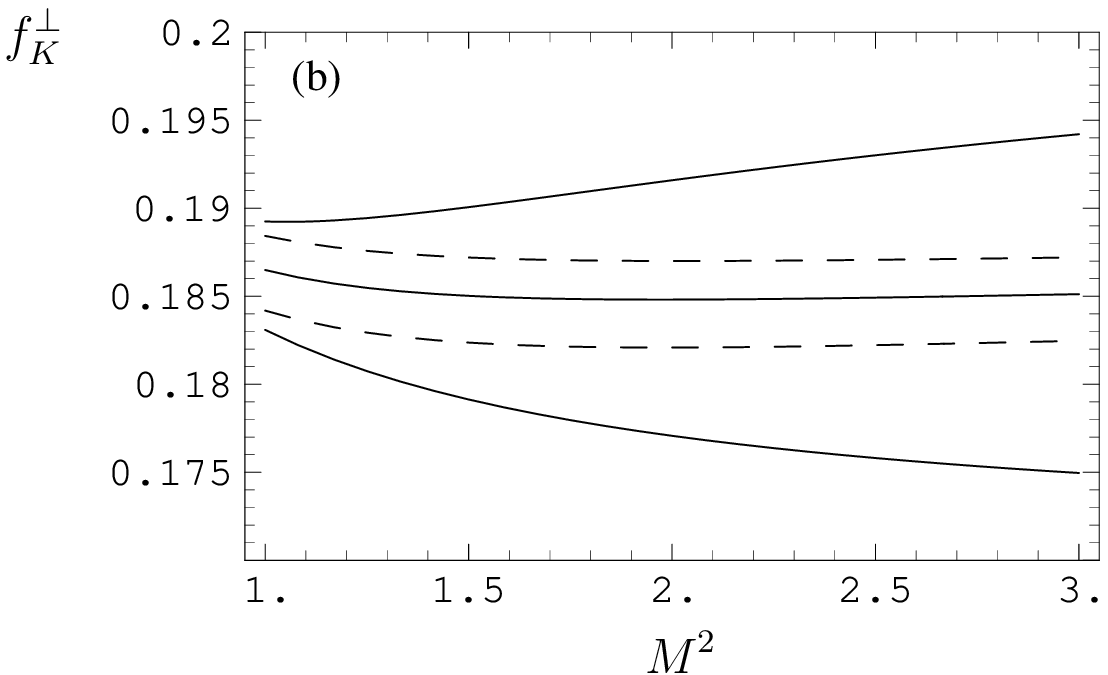}$$
\vskip-10pt
\caption[]{(a) $f_{K_1}^\perp(1\,{\rm GeV})$ from Eq.~(\ref{eq:pure}) 
as function of the Borel
  parameter $M^2$ for  $s_0=(2.7,2.9,3.1)\,{\rm GeV}^2$ (solid
  lines from bottom to top). 
(b) $f_K^\perp(1\,{\rm
    GeV})$ from Eq.~(\ref{eq:fKT}), using $f_{K_1}^\perp$ as
  input. Solid lines: $f_{K_1}^\perp=0.180\,$GeV and, from bottom to top,
   $s_0=(2.3,2.45,2.6)\,{\rm GeV}^2$. Dashed lines:
  $f_{K_1}^\perp=0.170\,$GeV (top) and 
  $f_{K_1}^\perp=0.190\,$GeV (bottom). Note that the optimum $s_0$ for
$f_{K_1}^\perp$ is larger than for $f_K^\perp$, in agreement with the
  resonance structure in the $1^+$ and $1^-$ channels.}\label{fig:2}
$$\epsfysize=0.19\textheight\epsffile{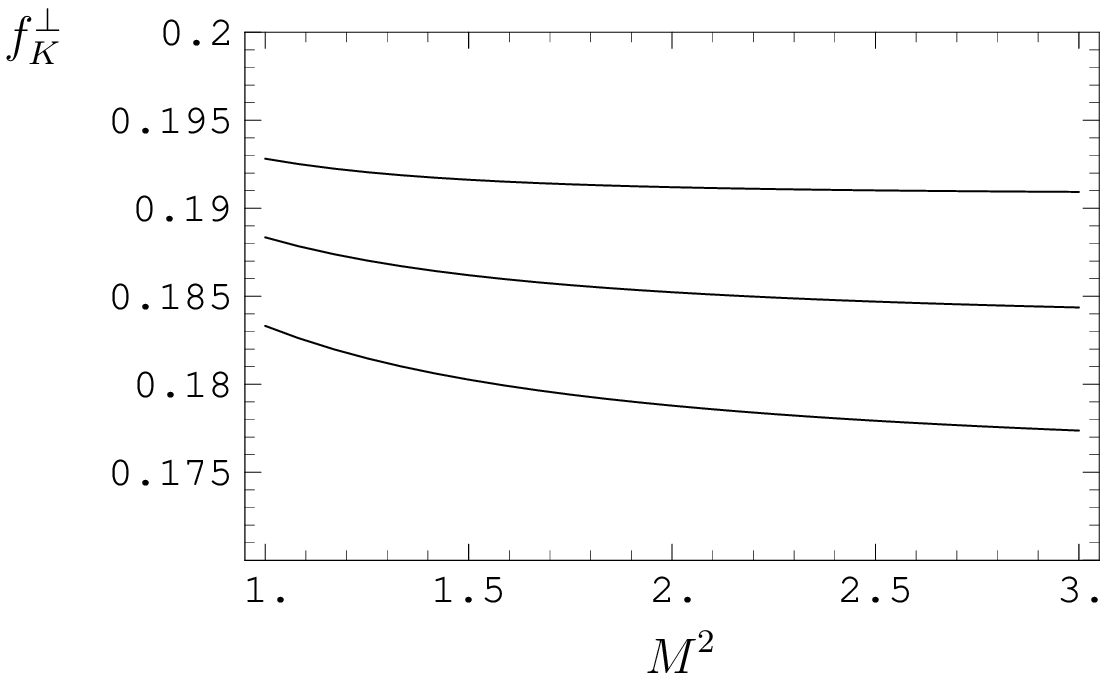}$$
\vskip-10pt
\caption[]{$f_K^\perp(1\,{\rm
    GeV})$ from Eq.~(\ref{eq:fKT}), setting $f_{K_1}^\perp=0$. From
    bottom to top: $s_0=(1.2,1.3,1.4)\,{\rm GeV}^2$.}\label{fig:extra}
\end{figure}

We determine $f_K^{(\parallel,\perp)}$ and
$a_1^{(\parallel,\perp)}(K^{(*)})$ 
from the diagonal correlation function
\begin{equation}\label{eq:3.4}
i\int d^4y e^{iqy} \langle 0 | T \bar q(y) \Gamma s(y) \bar s(0)\Gamma
[0,z] q(z) | 0\rangle,
\end{equation}
where $z_\mu$ is light-like and the Dirac structures $\Gamma$ are
given by
$$
K:\quad \Gamma = \hat{z}\gamma_5,\qquad 
K^{*}_\parallel:\quad \Gamma = \hat{z},\qquad 
K^{*}_\perp:\quad \Gamma = \sigma_{\mu\nu}z^\nu.
$$
The calculation with
 nonlocal operators is very convenient, as it allows one to calculate
 all moments in one go. 
Specifying for instance to $K^{*}_\parallel$, the sum rule reads
\begin{equation}\label{eq:x}
{\cal B_{\mbox{\scriptsize sub}}}\,\Pi^{\mbox{\scriptsize OPE}} =
(f_K^\parallel)^2\,e^{-m_{K^*}^2/M^2}\,\frac{1}{M^2}\int_0^1
du\,e^{i\ub\qz}\phi_K^\parallel(u),
\end{equation}
where also $\Pi^{\mbox{\scriptsize OPE}}$ is expressed as integral
over $u$, which naturally emerges as a Feynman parameter in the
calculation, and comes with the same weight function $\exp(i\ub\qz)$.
Sum rules for $f_K$ are obtained as the lowest order in an expansion in
$qz$, those for $a_1$ by effectively replacing
$$e^{i\ub\qz}\to \frac{5}{3}(u-\bar u).$$
For $f_K^{(\parallel)}$ we find the following sum rules:
$$
f_K^2 e^{-m_K^2/M^2}  = {\rm SR}_+,\qquad
(f_{K}^\parallel)^2 e^{-m_{K^*}^2/M^2}  = {\rm SR}_-,
$$
\begin{eqnarray}
\mbox{with~SR}_{\pm} & = & \frac{1}{4\pi^2}\int\limits_{m_s^2}^{s_0}
ds\,e^{-s/M^2} \,\frac{(s-m_s^2)^2 (s+2m_s^2)}{s^3} +
\frac{\alpha_s}{\pi}\, \frac{M^2}{4\pi^2}\left( 1 -
e^{-s_0/M^2}\right)\nonumber\\
&&{} +\frac{m_s\langle \bar s s\rangle}{M^2}\left(1+\frac{m_s^2}{3M^2} + 
\frac{13}{9}\,\frac{\alpha_s}{\pi}\right)+\frac{1}{12M^2}\,
\langle\frac{\alpha_s}{\pi}\,G^2\rangle 
\left( 1+ \frac{1}{3}\,\frac{m_s^2}{M^2}\right) \nonumber\\
&&{}  +\frac{4}{3}\,\frac{\alpha_s}{\pi} \, \frac{m_s\langle \bar q
  q\rangle}{M^2}\pm\frac{16\pi\alpha_s}{9M^4}\,
\langle \bar q q\rangle\langle \bar s s\rangle +
\frac{16\pi\alpha_s}{81M^4}\,\left( \langle \bar q q\rangle^2 +
\langle \bar s s\rangle^2 \right).
\end{eqnarray}
This sum rule was already given in Ref.~\cite{L1}, apart from the radiative and
mass-corrections to the quark-condensate contribution which are new. 
The values of $f_K$ and $f_K^\parallel$ obtained from these sum rules
are shown in Fig.~\ref{fig:1}, evaluating all scale-dependent
quantities at the scale $\mu=1\,$GeV.
The sum rule results agree very well with the experimental values \cite{PDG}
$$f_K = (0.160\pm 0.002)\,{\rm GeV},\qquad 
f_K^\parallel = (0.217\pm 0.005)\,{\rm GeV}. 
$$
For $f_K^\perp$ the situation is slightly more subtle as the
correlation function (\ref{eq:3.4}) contains contributions not only
from the vector meson $K^*$, but also from the axial-vector meson
$K_1$. The same situation occurs for $\rho$ and $b_1(1235)$. In
Ref.~\cite{BB96}, where QCD sum rules for $f_\rho^\perp$ were
studied, it was argued that one can either explicitly 
include the contribution of $b_1$ in the hadronic parametrisation of
this ``mixed-parity''
sum rule and use a suitably large value of the continuum threshold
$s_0\approx 2.1\,{\rm GeV}^2$ or include it in the continuum and use a
smaller value $s_0\approx 1.0\,{\rm GeV}^2$. Both procedures yield a
stable sum rule and $f_\rho^\perp(1\,{\rm GeV})\approx 160\,$MeV. For
$K^*$, however, the mixed-parity 
sum rule without $K_1$
does not display a stable plateau in $M^2$, which means one has to include
the contribution of $K_1$ explicitly.\footnote{$K_1$ also contributes
  to the sum rule for $f_K$, but  
can be safely absorbed into the
  continuum, as $m_{K_1}\gg m_K$.} 
There are actually two strange axial-vector 
mesons, $K_1(1270)$ and $K_1(1400)$, which are usually
interpreted as mixture of a ${}^3 P_1$ state, the $K_a$, and a
${}^1 P_1$ state, the $K_b$ \cite{Suzuki,Goldman}:
\begin{eqnarray*}
K_1(1270) & = & K_a \cos \theta_K - K_b \sin \theta_K,\\
K_1(1400) & = & K_a \sin \theta_K + K_b \cos \theta_K.
\end{eqnarray*}
The results of Refs.~\cite{Suzuki,Goldman} indicate that the system is
close to ideal mixing, i.e.\ $\theta_K \approx 45^\circ$. To the
accuracy needed in our sum rules it is then sufficient to replace the
two resonances by one effective one with the mass $m_{K_1} = 1.34\,$GeV
\cite{Goldman}. 
The mixed-parity sum rule obtained from the correlation function
(\ref{eq:3.4}) now reads:
\begin{eqnarray}
\lefteqn{(f_{K}^\perp)^2 e^{-m_{K^*}^2/M^2} +(f_{K_1}^\perp)^2
 e^{-m_{K_1}^2/M^2}  =}\hspace*{0cm}\nonumber\\
& & \frac{1}{4\pi^2}\int\limits_{m_s^2}^{s_0}
ds\,e^{-s/M^2} \,\frac{(s-m_s^2)^2 (s+2m_s^2)}{s^3} +
\frac{1}{4\pi^2}\int\limits_{0}^{s_0}
ds\,e^{-s/M^2} \,\frac{\alpha_s}{\pi}\left( \frac{7}{9} +
\frac{2}{3}\,\ln \frac{s}{\mu^2}\right)\nonumber\\
&&{}+\frac{m_s\langle \bar s
  s\rangle}{M^2}\left\{1+\frac{m_s^2}{3M^2}+
\frac{\alpha_s}{\pi}\left(-\frac{19}{9} + \frac{2}{3}
\left[ 1-\gamma_E + \ln\,\frac{M^2}{\mu^2} +
  \frac{M^2}{s_0}\,e^{-s_0/M^2} + {\rm Ei}\left(-\frac{s_0}{M^2}\right)\right]
\right)\right\}\nonumber
\end{eqnarray}
\begin{eqnarray}
&&{} -\frac{1}{12M^2}\,\langle\frac{\alpha_s}{\pi}\,G^2\rangle 
\left\{ 1-\frac{2m_s^2}{M^2}\left( \frac{7}{6}-\gamma_E + {\rm
  Ei}\left(
-\frac{s_0}{M^2}\right) - \ln\,\frac{m_s^2}{M^2} +
\frac{M^2}{s_0}\left( 1 - \frac{M^2}{s_0}\right) e^{-s_0/M^2}
\right)\right\} \nonumber\\
&&{} -\frac{1}{3M^4}\,m_s\langle \bar s\sigma gGs\rangle -
\frac{32\pi\alpha_s}{81M^4}\,\left( \langle \bar q q\rangle^2 +
\langle \bar s s\rangle^2 \right).\label{eq:fKT}
\end{eqnarray}
The value of $f_{K_1}^\perp$ itself can be obtained from a ``pure-parity''
sum rule which can be extracted from the correlation function
(\ref{eq:3.4}) leaving one index uncontracted; explicit expressions are
given in Ref.~\cite{L1}. As is well-known, 
the diadvantage of pure-parity sum rules is that they
come with a higher mass-dimension which increases the dependence of
the result on the continuum model, which is however acceptable as long as as we
only need an estimate of $f_{K_1}^\perp$  for use in
Eq.~(\ref{eq:fKT}).
The pure-parity sum rule for $f_{K_1}^\perp$, including SU(3)-breaking
corrections, reads \cite{L1}:
\begin{eqnarray}
\lefteqn{(f_{K_1}^\perp)^2 m_{K_1}^2 e^{-m_{K_1}^2/M^2} = 
\frac{1}{8\pi^2}\int\limits_{m_s^2}^{s_0}
ds\,e^{-s/M^2} \,\frac{(s-m_s^2)^2 (s+2m_s^2)}{s^2}}\nonumber\\
&&{} +
\frac{1}{8\pi^2}\int\limits_{0}^{s_0}
ds\,s e^{-s/M^2} \,\frac{\alpha_s}{\pi}\left( \frac{7}{9} +
\frac{2}{3}\,\ln \frac{s}{\mu^2}\right) +
\frac{32\pi\alpha_s}{81M^2}\,\left( \langle \bar q q\rangle^2 +
\langle \bar s s\rangle^2 \right) -
\frac{16\pi\alpha_s}{9M^2}\,\langle \bar q q\rangle\langle \bar s
s\rangle\nonumber\\
&&{}+m_s\langle\bar q q \rangle -\frac{m_s\langle \bar s
  s\rangle}{2} +\frac{1}{24M^2}\,\langle\frac{\alpha_s}{\pi}\,G^2\rangle 
+\frac{1}{3M^2}\,m_s\langle \bar s\sigma gGs\rangle
-\frac{1}{6M^2}\,m_s\langle \bar q\sigma gGq\rangle.\label{eq:pure}
\end{eqnarray}
Again all scale-dependent quantities are evaluated at $\mu=1\,$GeV. 
The results for $f_{K_1}^\perp$ are shown in Fig.~\ref{fig:2}(a), from
which we conclude
\begin{equation}\label{grunz}
f_{K_1}^\perp(1\,{\rm GeV}) = (0.185\pm 0.010)\,{\rm GeV}.
\end{equation}
This result is slightly smaller than, but still in agreement with, the
one obtained in Ref.~\cite{Yang}.
Using (\ref{grunz}) as input in (\ref{eq:fKT}), we obtain
the values for $f_K^\perp(1\,{\rm GeV})$ shown in
Fig.~\ref{fig:2}(b), yielding
\begin{equation}\label{eq:resfKT}
f_{K}^\perp(1\,{\rm GeV}) = (0.185\pm 0.010)\,{\rm GeV}.
\end{equation}
The value of $f_{K}^\perp$ at different scales can be obtained from
the leading-order renormalisation-group improved relation
$$
f_{K}^\perp(\mu) = f_{K}^\perp(1\,{\rm GeV})
\left(\frac{\alpha_s(\mu)}{\alpha_s(1\,{\rm GeV})}\right)^{4/(3\beta_0)}.
$$
The result (\ref{eq:resfKT}) has to be compared with $(0.170\pm
0.010)\,{\rm GeV}$ quoted in Ref.~\cite{BZvector}. The main difference
is that the sum rule (\ref{eq:fKT}) includes, in addition to the new
terms in $\alpha_s m_s \squark$ and $m_s^2 \gluon$, in particular 
the contribution of $f_{K_1}^\perp$, which allows one to obtain a
stable plateau in $M^2$. Once we know the value of $f_{K}^\perp$, we
can now determine the continuum threshold to be used when $K_1$ is
included in the continuum, which will be relevant for the
determination of $a_1^\perp(K^*)$. Fig.~\ref{fig:extra} shows that
$s_0=(1.3\pm 0.1)\,{\rm GeV}^2$ is the appropriate value to use if the
contribution of $K_1$ is not  made explicit. 

$f_K^\perp$ given in (\ref{eq:resfKT}) also compares
favourably with the average result from lattice calculations \cite{lattbec}:
$$ \frac{f_K^\perp(2\,{\rm GeV})}{f_K^\parallel} = 0.76\pm 0.03
~\leftrightarrow~ f_{K}^\perp(1\,{\rm GeV}) = (0.178\pm 0.005)\,{\rm
  GeV}.
$$

We are now in a position to analyse the sum rules for
$a_1$. The sum rule
for $a_1(K)$ agrees with the one obtained in Ref.~\cite{alex}, that
for $a_1^\parallel(K^*)$ is new. We find
\begin{eqnarray}
\lefteqn{a_1(K) f_K^2 e^{-m_K^2/M^2}  =  {\rm SR}_{1,+},\qquad
a_1^{\parallel}(K^*)(f_{K}^\parallel)^2 e^{-m_{K^*}^2/M^2}  = 
{\rm SR}_{1,-},}\nonumber\\
\lefteqn{\mbox{with~SR}_{1,\pm} = 
\frac{5}{4\pi^2}\,m_s^4\int\limits_{m_s^2}^{s_0}
ds\,e^{-s/M^2} \,\frac{(s-m_s^2)^2}{s^4}}\nonumber\\
\hskip-10pt&&
{}+\frac{5m_s^2}{18M^4}\,\langle\frac{\alpha_s}{\pi}\,G^2\rangle 
\left( -\frac{1}{2}+ \gamma_E - {\rm
  Ei}\left(-\frac{s_0}{M^2}\right)+ \ln\,\frac{m_s^2}{M^2} + 
\frac{M^2}{s_0}\left( \frac{M^2}{s_0} -1\right) e^{-s_0/M^2}\right)\nonumber\\
&&{}
-\frac{5}{3}\,\frac{m_s\langle \bar s
  s\rangle}{M^2}\left\{1+ \frac{m_s^2}{M^2} + 
\frac{\alpha_s}{\pi}\left[ -\frac{124}{27} +
\frac{8}{9} \left(1-\gamma_E + \ln\,\frac{M^2}{\mu^2} +
\frac{M^2}{s_0}\,e^{-s_0/M^2} + {\rm Ei}\left(-\frac{s_0}{M^2}\right)
\right)\right]\right\}  \nonumber\\
&&{}\mp 
\frac{20}{27}\,\frac{\alpha_s}{\pi} \, \frac{m_s\langle \bar q
 q\rangle}{M^2} + \frac{5}{9}\, \frac{m_s\langle\bar s \sigma g Gs\rangle}{M^4}
+\frac{80\pi\alpha_s}{81M^4}\,\left( \langle \bar q q\rangle^2 -
\langle \bar s s\rangle^2 \right).\label{eq:SRa1}
\end{eqnarray}
\begin{figure}[p]
$$\epsfysize=0.19\textheight\epsffile{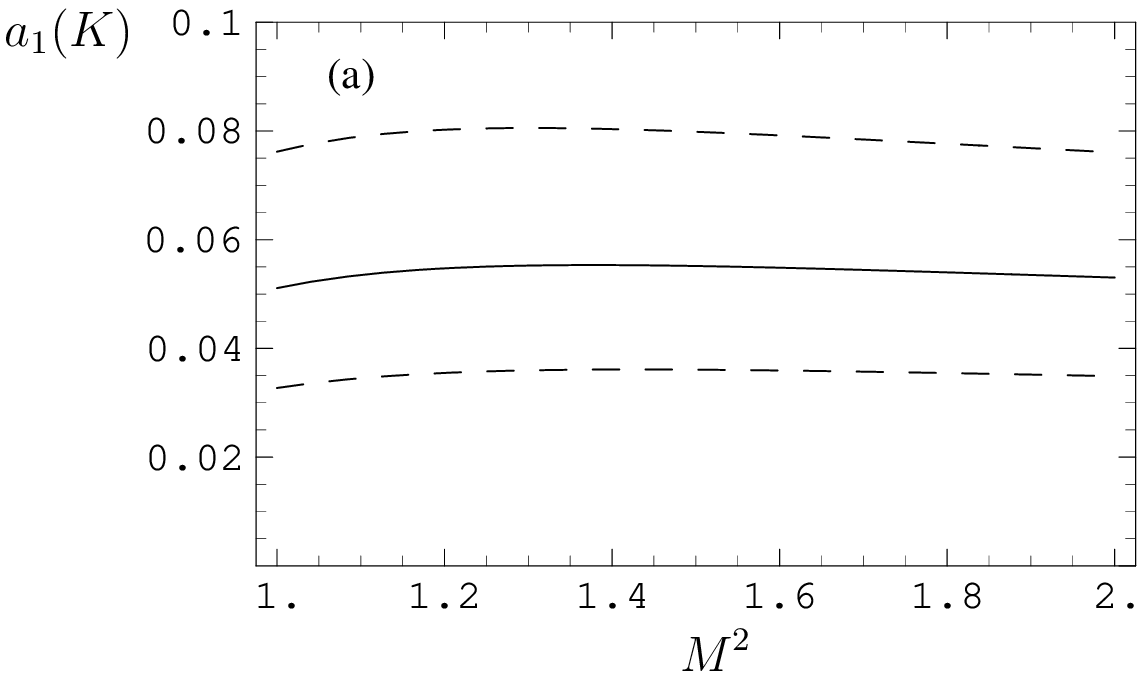}\qquad
\epsfysize=0.19\textheight\epsffile{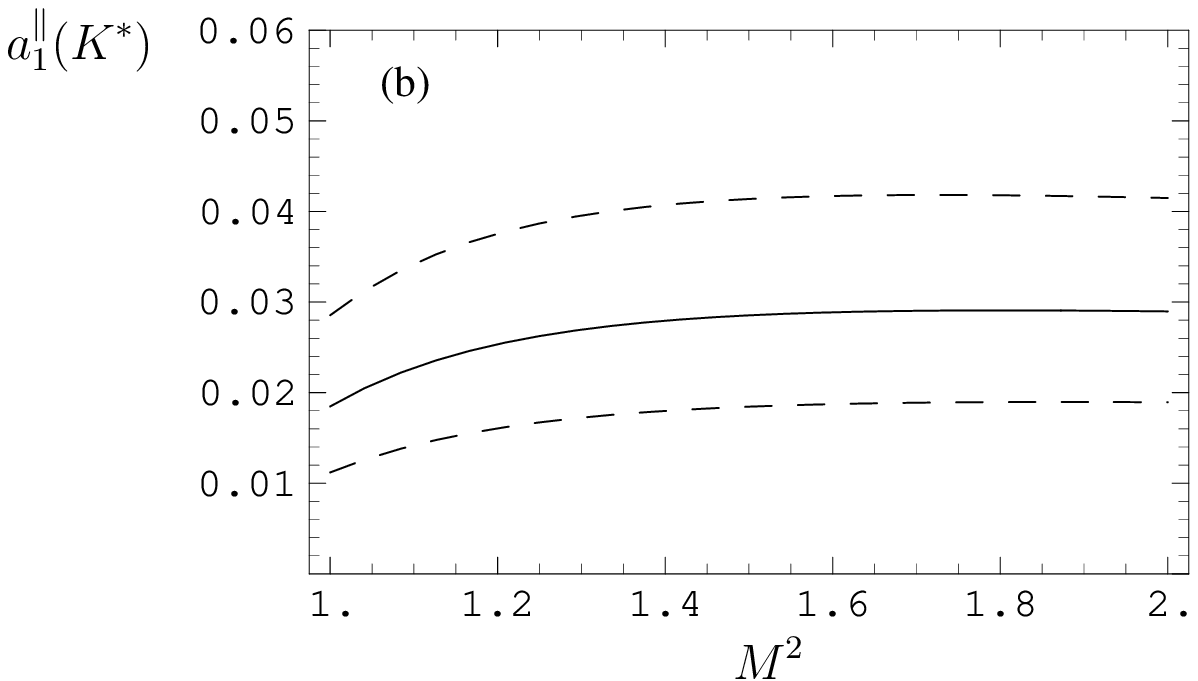}$$
\caption[]{(a) $a_1(K)$ as function of the Borel parameter
  $M^2$ for $s_0=1.1\,{\rm GeV}^2$ and $\mu=1\,$GeV. 
Solid line: central values of
  input parameters, dashed lines: variation of $a_1(K)$ within the
  allowed range of input parameters. (b): Same for
  $a_1^{\parallel}(K^*)$ with $s_0 = 1.7\,{\rm GeV}^2$.}\label{fig:3}
$$\epsfysize=0.19\textheight\epsffile{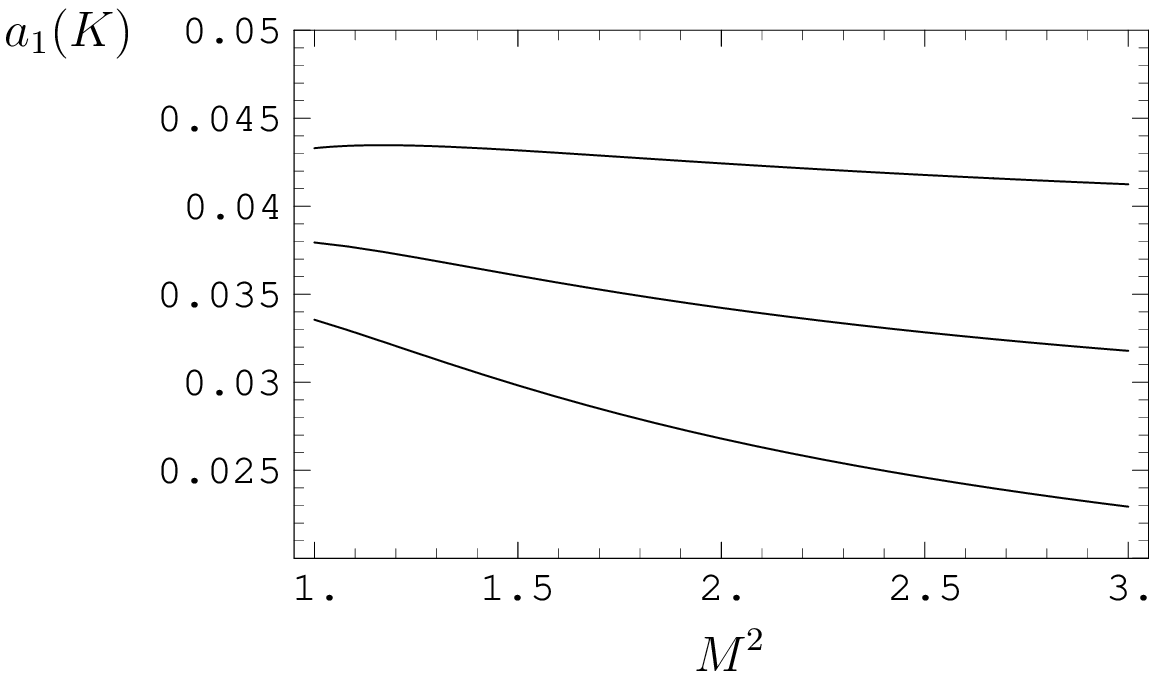}$$
\vskip-10pt
\caption[]{$a_1(K)$ from the (tree-level) 
nondiagonal sum rule (6.27) in
Ref.~\cite{CZreport} after the correction of a sign-mistake in the
perturbative contribution. $s_0 = (1.0,1.2,1.4)\,{\rm GeV}^2$ (from
top to bottom).}\label{fig:4}
$$\epsfysize=0.19\textheight\epsffile{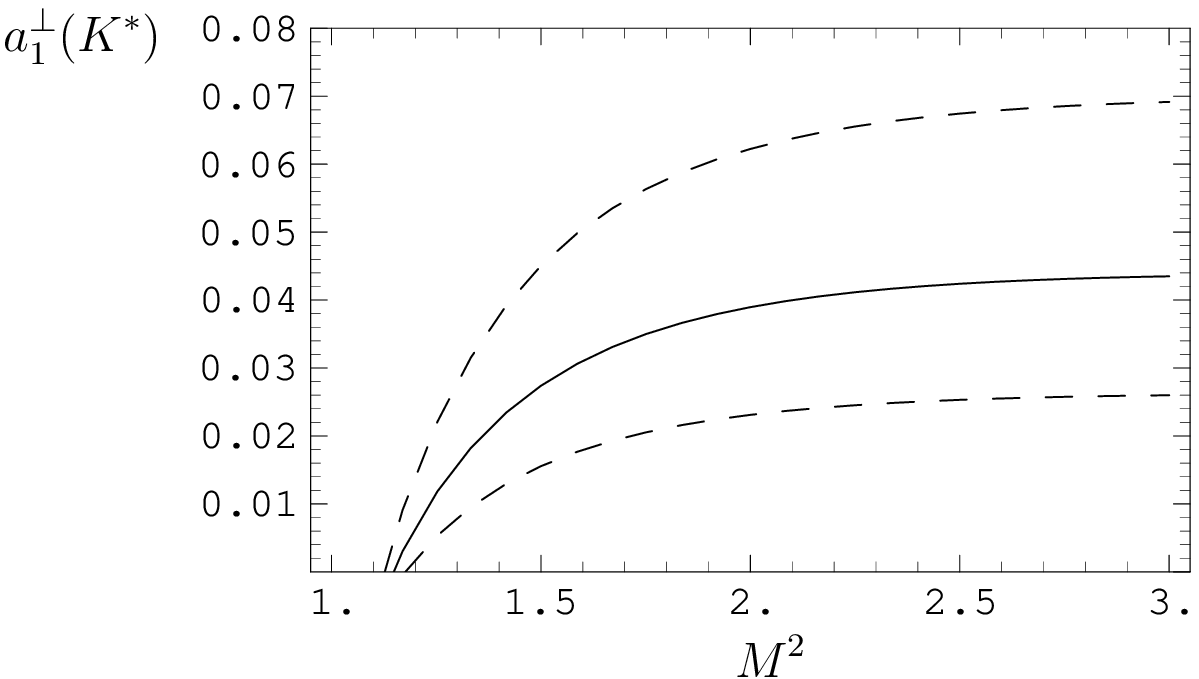}$$
\vskip-10pt
\caption[]{$a_1^\perp(K^*,1\,{\rm GeV})$  as function of the Borel parameter
  $M^2$ for $\mu=1\,$GeV. Solid line: 
central values of input parameters, dashed lines: 
variation of $a_1^\perp(K^*)$ within the
  allowed range of input parameters and $s_0 = (1.3\pm 0.1)\,{\rm
  GeV}^2$.}\label{fig:5}
\end{figure}
The sum rules are plotted in Fig.~\ref{fig:3}; the dominant terms are
those in the quark and mixed condensates. As the perturbative
contribution is $\sim O(m_s^2)$ and hence small, the sum rules are not
stable in $M^2$, so that we choose to 
evaluate the sum rules using the optimum values 
of $s_0$ as determined
from the calculation of the decay constants. As expected, the
dependence of $a_1$ on SU(3)-breaking parameters, in particular the
precise value of $m_s$, is much stronger than for the decay constants
$f_K$. From Fig.~\ref{fig:3} we read off the following results for
$a_1$:
\begin{equation}\label{eq:resa1}
a_1(K,1\,{\rm GeV}) = 0.050\pm 0.025,\qquad a_1^\parallel(K^*,1\,{\rm
  GeV}) = 0.025\pm 0.015.
\end{equation}
The value for $a_1(K,1\,{\rm GeV})$ agrees with the one obtained in
Ref.~\cite{alex}, although our uncertainty is slightly
larger. $a_1^\parallel(K^*)$ is smaller than $a_1(K)$, which follows
from the fact that the right-hand sides of the sum rules
(\ref{eq:SRa1}) are essentially  the same, except for the values of
$s_0$, and one term which gives
a positive contribution to $a_1(K)$, but a negative to
$a_1^\parallel(K^*)$. Since $f^\parallel_K>f_K$, and the sensitivity
of the sum rule on $s_0$ is small, one clearly expects
$a_1^\parallel(K^*)< a_1(K)$. 
Both
results, however, markedly disagree with those obtained in
Refs.~\cite{CZreport,elena}. 
In Sec.~\ref{sec:1}  we have already mentioned the
reasons for this discrepancy: in Ref.~\cite{elena} a different set of
sum rules, nondiagonal sum rules and a chirally odd correlation
function, were used which exhibit large cancellations between the
dominant terms. As discussed in Ref.~\cite{alex,lenz}, these sum rules are
numerically not reliable. Eq.~(\ref{eq:SRa1}) is free of such
cancellations and hence expected to be more reliable. On the other
hand, $a_1(K)$ and $a_1^\parallel(K^*)$ 
in (\ref{eq:resa1}) are also smaller than the orginal
results of Chernyak and Zhitnitsky \cite{CZreport}. This is due to the
fact that in the corresponding nondiagonal sum rule Eq.~(6.27) in
Ref.~\cite{CZreport}, which contains no radiative corrections,
the perturbative term has the wrong
sign. Correcting the sign, and using the standard values of input
parameters from Tab.~\ref{tab:cond}, we obtain the results shown in
Fig.~\ref{fig:4}, which are remarkably close to our result
(\ref{eq:resa1}) from the diagonal sum rule. Unfortunately, once radiative
corrections are included in the nondiagonal sum rule, the agreement
with (\ref{eq:resa1}) is lost and one is back to the results obtained
in Ref.~\cite{elena} with negative $a_1$. 

Let us finally turn to $a_1^\perp(K^*)$. As with the decay constant,
the sum rule obtained from the correlation function (\ref{eq:3.4})
contains contributions from the $K_1$, which in principle need to be 
subtracted. Based on the experience with the sum rule for
$f_{K}^\perp$, however, we decide to include these contributions in
the continuum and use the continuum threshold $s_0 = (1.3\pm
0.1)\,{\rm GeV}^2$ determined from Fig.~\ref{fig:extra}. The sum rule
for $a_1^\perp(K^*)$ reads
\begin{eqnarray}
\lefteqn{a_1^{\perp}(K^*)(f_{K}^\perp)^2 e^{-m_{K^*}^2/M^2} =  
\frac{5}{4\pi^2}\,m_s^4\int\limits_{m_s^2}^{s_0}
ds\,e^{-s/M^2} \,\frac{(s-m_s^2)^2}{s^4}+ \frac{10}{9}\, 
\frac{m_s\langle\bar s \sigma g Gs\rangle}{M^4}} \nonumber\\[-7pt]
&&{}+\frac{5m_s^2}{9M^4}\,\langle\frac{\alpha_s}{\pi}\,G^2\rangle 
\left( \frac{1}{4}+ \gamma_E - {\rm
  Ei}\left(-\frac{s_0}{M^2}\right)+ \ln\,\frac{m_s^2}{M^2} + 
\frac{M^2}{s_0}\left( \frac{M^2}{s_0} -1\right) e^{-s_0/M^2}\right)\nonumber\\
&&{}-\frac{5}{3}\,\frac{m_s\langle \bar s
  s\rangle}{M^2}\left\{1+ \frac{m_s^2}{M^2} + 
\frac{\alpha_s}{\pi}\left[ -\frac{46}{9} +
\frac{4}{3} \left(1-\gamma_E + \ln\,\frac{M^2}{\mu^2} +
\frac{M^2}{s_0}\,e^{-s_0/M^2} + {\rm Ei}\left(-\frac{s_0}{M^2}\right)
\right)\right]\right\}\nonumber\\[-10pt]\label{eq:a1perp}
\end{eqnarray}
and is plotted in Fig.~\ref{fig:5}. We find
\begin{equation}
a_1^\perp(K^*,1\,{\rm GeV}) = 0.04\pm 0.03.
\end{equation}

\section{Summary and Conclusions}\label{sec:4}

In this letter we have calculated the decay constants $f_K$ and
$f_K^{\parallel,\perp}$ of the $K$ and $K^*$ meson, respectively,
and the first Gegenbauer moments of the leading-twist DAs, $a_1(K)$
and $a_1^{\parallel,\perp}(K^*)$. We find values for $f_K$ and
$f_K^\parallel$ in agreement with experiment and
\begin{eqnarray*}
&&f_{K}^\perp(1\,{\rm GeV}) =  (0.185\pm 0.010)\,{\rm GeV},\quad
a_1(K,1\,{\rm GeV}) = 0.050\pm 0.025,\\
&&a_1^\parallel(K^*,1\,{\rm
  GeV}) = 0.025\pm 0.015,\quad\quad
~\,a_1^\perp(K^*,1\,{\rm GeV}) = 0.04\pm 0.03.
\end{eqnarray*}
The value for $f_K^\perp$ agrees, within uncertainties, with that
found from lattice calculations \cite{lattbec}. The value of $a_1(K)$
agrees with that found in Ref.~\cite{alex}, using the same
method. $a_1(K)$ and $a_1^{\parallel,\perp}(K^*)$ disagree with the
negative values found in Ref.~\cite{elena} which is due to numerical
instabilities of the nondiagonal sum rules used in that
paper. $a_1(K)$ and $a_1^\parallel(K^*)$ are also smaller, by roughly
a factor two, than the results obtained in Ref.~\cite{lenz} from exact
operator relations between $a_1(K)$ and $a_1^{\parallel}(K^*)$ and
quark-quark-gluon matrix elements. Whereas it seems rather unlikely
that an increase in the accuracy of the sum rules for
$a_1$, Eqs.~(\ref{eq:SRa1}), (\ref{eq:a1perp}), by
including corrections in $m_s^2\alpha_s$ and $m_s\alpha_s\mixed$
will change the results for $a_1$ by a factor
two, the situation may be different with the sum rules for the
quark-quark-gluon matrix elements employed in Ref.~\cite{lenz}, which
only contain terms in $\sim m_s$. We plan to come back to this question in
a separate publication.

\subsection*{Acknowledgements}
R.Z.\ is supported in part by the EU-RTN Programme, Contract No.\
HPEN-CT-2002-00311, ``EURIDICE''. 

\appendix
\setcounter{equation}{0}
\renewcommand{\theequation}{A.\arabic{equation}}

\section{Some Details of the Calculation}

In calculating the contribution of the strange quark condensate, one
has to include the first nonlocal term in the expansion of the
quark condensate:
\begin{eqnarray}
\langle 0 | \bar s_{i\alpha}(x) s_{j\beta}(y) | 0 \rangle & = &
\frac{1}{12}\,\langle \bar s s \rangle \delta_{ij} \left(
\delta_{\alpha\beta} + \frac{i}{D}\,m_s (x_\kappa - y_\kappa)
\left(\gamma^\kappa\right)_{\beta\alpha}\right)\nonumber\\
& = & \frac{1}{12}\,\langle \bar s s \rangle \delta_{ij} \left.\left(
\delta_{\alpha\beta} + \frac{1}{D}\,m_s \frac{\partial}{\partial Q_\kappa}
\left(\gamma^\kappa\right)_{\beta\alpha}\right)e^{-i
  Q(y-x)}\right|_{Q=0},\label{eq:app1}
\end{eqnarray}
where $i,j$ are colour, $\alpha,\beta$ spinor indices, and
$D$ is the
number of dimensions; $Q$ is an
auxiliary momentum. It is the second term in (\ref{eq:app1}) that
causes a slight complication in the calculation of radiative
corrections in the form of finite counter terms. Their origin is
twofold: first, the factor $1/D$ induces $O(\epsilon)$ contributions
at tree-level which cause finite counter terms upon renormalisation.
Second, if the derivative in $Q_\kappa$ yields a term
$\gamma_\kappa$ in the trace, the contraction over $\kappa$ can also
yield finite terms in the counter term, which indeed happens for the
vertex correction diagrams. 
It appears that these finite counter terms
have been missed by the authors of Ref.~\cite{alex}, for we
reproduce the results in their appendix using
(\ref{eq:app1}) and dropping just the divergent terms.


\begin{thebibliography}{99}

\bibitem{exclusive}
V.L.\ Chernyak and A.R.\ Zhitnitsky,
JETP Lett.\  {\bf 25} (1977) 510
[Pisma Zh.\ Eksp.\ Teor.\ Fiz.\  {\bf 25} (1977) 544];
Sov.\ J.\ Nucl.\ Phys.\  {\bf 31} (1980) 544
[Yad.\ Fiz.\  {\bf 31} (1980) 1053];\\
A.V.\ Efremov and A.V.\ Radyushkin,
Phys.\ Lett.\ B {\bf 94} (1980) 245;
Theor.\ Math.\ Phys.\  {\bf 42} (1980) 97
[Teor.\ Mat.\ Fiz.\  {\bf 42} (1980) 147];\\
G.P.\ Lepage and S.J.\ Brodsky,
Phys.\ Lett.\ B {\bf 87} (1979) 359;
Phys.\ Rev.\ D {\bf 22} (1980) 2157;\\
V.L.\ Chernyak, A.R.\ Zhitnitsky and V.G.\ Serbo,
JETP Lett.\  {\bf 26} (1977) 594
[Pisma Zh.\ Eksp.\ Teor.\ Fiz.\  {\bf 26} (1977) 760];
Sov.\ J.\ Nucl.\ Phys.\  {\bf 31} (1980) 552
[Yad.\ Fiz.\  {\bf 31} (1980) 1069].

\bibitem{BBNS}
M. Beneke {\em et al.},
Phys.\ Rev.\ Lett.\  {\bf 83} (1999) 1914
[arXiv:hep-ph/9905312];
Nucl.\ Phys.\ B {\bf 591} (2000) 313
[arXiv:hep-ph/0006124];
Nucl.\ Phys.\ B {\bf 606} (2001) 245
[arXiv:hep-ph/0104110].

\bibitem{Russians}
V.~L.~Chernyak, A.~R.~Zhitnitsky and I.~R.~Zhitnitsky,
Nucl.\ Phys.\ B {\bf 204} (1982) 477
[Erratum-ibid.\ B {\bf 214} (1983) 547];
Sov.\ J.\ Nucl.\ Phys.\  {\bf 38} (1983) 775
[Yad.\ Fiz.\  {\bf 38} (1983) 1277].

\bibitem{CZreport} 
V.~L.~Chernyak and A.~R.~Zhitnitsky,
Phys.\ Rept.\  {\bf 112} (1984) 173.

\bibitem{elena} P.~Ball and M.~Boglione,
  Phys.\ Rev.\ D {\bf 68}, 094006 (2003)
  [arXiv:hep-ph/0307337].

\bibitem{alex}   A.~Khodjamirian, T.~Mannel and M.~Melcher,
  Phys.\ Rev.\ D {\bf 70} (2004) 094002
  [arXiv:hep-ph/0407226].

\bibitem{lenz}  V.~M.~Braun and A.~Lenz,
  Phys.\ Rev.\ D {\bf 70}, 074020 (2004)
  [arXiv:hep-ph/0407282].

\bibitem{lattbec}   D.~Becirevic, V.~Lubicz, F.~Mescia and C.~Tarantino,
  JHEP {\bf 0305}, 007 (2003)
  [arXiv:hep-lat/0301020];\\
V.~M.~Braun {\it et al.},
Phys.\ Rev.\ D {\bf 68} (2003) 054501
[arXiv:hep-lat/0306006].

\bibitem{BZvector}  P.~Ball and R.~Zwicky,
  Phys.\ Rev.\ D {\bf 71} (2005) 014029
  [arXiv:hep-ph/0412079].

\bibitem{SVZ}
M.A.~Shifman, A.I.~Vainshtein and V.I.~Zakharov,
Nucl.\ Phys.\ B {\bf 147} (1979) 385;
Nucl.\ Phys.\ B {\bf 147} (1979) 448.

\bibitem{colalex}
P.~Colangelo and A.~Khodjamirian,
arXiv:hep-ph/0010175.

\bibitem{FFs}
P.~Ball,
JHEP {\bf 9809} (1998) 005
[arXiv:hep-ph/9802394];\\
P.~Ball and V.~M.~Braun,
Phys.\ Rev.\ D {\bf 58} (1998) 094016
[arXiv:hep-ph/9805422];\\
 P.~Ball and R.~Zwicky,
  JHEP {\bf 0110} (2001) 019
  [arXiv:hep-ph/0110115];
Phys.\ Rev.\ D {\bf 71} (2005) 014015
  [arXiv:hep-ph/0406232];
Phys.\ Lett.\ B {\bf 625}, 225 (2005)
[arXiv:hep-ph/0507076].

\bibitem{Wittig}
S.~Hashimoto,
arXiv:hep-ph/0411126.

\bibitem{L1}
J.~Govaerts {\it et al.},
Nucl.\ Phys.\ B {\bf 283} (1987) 706.

\bibitem{PDG}
S.~Eidelman {\it et al.}  [Particle Data Group Collaboration],
Phys.\ Lett.\ B {\bf 592} (2004) 1.

\bibitem{BB96}   P.~Ball and V.~M.~Braun,
  Phys.\ Rev.\ D {\bf 54}, 2182 (1996)
  [arXiv:hep-ph/9602323].

\bibitem{Suzuki}
M.~Suzuki,
Phys.\ Rev.\ D {\bf 47} (1993) 1252.

\bibitem{Goldman}
L.~Burakovsky and T.~Goldman,
Phys.\ Rev.\ D {\bf 57} (1998) 2879
[arXiv:hep-ph/9703271].

\bibitem{Yang}
K.~C.~Yang,
arXiv:hep-ph/0509337.
\end{thebibliography}
\end{document}